\documentclass[prd,preprint,
superscriptaddress,showpacs,nofootinbib,%
tightenlines
]{revtex4}
\usepackage{epsfig}
\usepackage{amssymb}

\newcommand{\ben}{\begin{displaymath}}
\newcommand{\een}{\end{displaymath}}
\newcommand{\be}{\begin{equation}}
\newcommand{\ee}{\end{equation}}
\newcommand{\bea}{\begin{eqnarray}}
\newcommand{\eea}{\end{eqnarray}}
\begin{document}
\title{Why gauge symmetry?}
\author{J.~Gegelia}
\affiliation
{Institut f\"ur Theoretische Physik II, Ruhr-Universit\"at Bochum, 44780 Bochum, Germany}
\affiliation{
Tbilisi State University, 0186 Tbilisi,
Georgia}
\date{1 June, 2012}
\begin{abstract}
It is argued that the Weinberg-Salam model is the way it is because the most general self-consistent effective field theory of massive vector bosons interacting with fermions and photons at leading order coincides with the Weinberg-Salam model in unitary gauge where the scalar field is replaced by its vacuum expectation value. To support this argument the most general Lorentz-invariant
effective Lagrangian of massive vector bosons coupled to massless
fermions is considered. Restrictions imposed on the interaction terms
following from the consistency with the constraints of the second class and the perturbative renormalizability
in the sense of effective field theories is analyzed. It is shown that the leading order effective Lagrangian
containing interaction terms with dimensionless coupling constants coincides with the leading order effective
Lagrangian of the locally invariant Yang-Mills theory up to globally invariant mass term of the vector bosons.
Including the fermion masses and mixings and the interaction with the electromagnetic field leads to an effective field theory which at leading order looks like as if it was an $SU(2)\times U(1)$ gauge invariant theory with spontaneous symmetry breaking in unitary gauge with the scalar field replaced by its vacuum expectation value.

\end{abstract}



\pacs{04.60.Ds, 11.10.Gh, 03.70.+k, \\
Keywords: Effective field theory; Quantization; Constraints; Renormalization;  Electro-weak interaction}

\maketitle

\section{Introduction}

	Due to its impressive success in describing the experimental data, the
standard model (SM) is widely accepted as an established consistent theory of
strong, electromagnetic and weak interactions.  The modern
point of view is to think of the SM as an effective field theory (EFT),
``low-energy approximation to a deeper theory that may not even be
a field theory, but something different like a string theory''
\cite{Weinberg:mt}. The effective Lagrangian consists of an
infinite number of terms. However the coupling constants of
non-renormalizable interactions are suppressed by powers of a large scale, so that
their contributions in physical quantities are negligible for energies much
lower than the large scale.
	Renormalizability in the sense of fundamental theory is replaced by the
renormalizability in the sense of EFT, i.e. that all divergences can be absorbed by renormalizing an infinite number of parameters of the effective Lagrangian.

	In modern approach to quantum field theories one usually takes gauge invariance as the
starting point. However, in trying to understand ''why this theory takes the form it does, and why in this
form it does such a good job of describing the real world'' \cite{Weinberg:mt} it is difficult to justify the principle of local gauge invariance.
In particular, the electromagnetic and gravitational forces are long-range, therefore they must be described by gauge
theories \cite{Weinberg:mt}. On the other hand, as the weak interaction is mediated by massive particles,
it is not clear why should it be described by a gauge theory. ''It isn't any good just to present the formalism and say that it agrees with experiment -- you have to explain ... why this is the way the world is. After all, this is our aim in physics, not just to describe nature, but
to explain nature'' \cite{Weinberg:1996kw}.
	Despite the great success of the Weinberg-Salam (WS) model \cite{Weinberg:1967tq,Salam:1968rm}, it is unclear why is
this model the way it is. If one includes scalars and introduces masses
via Higgs mechanism in the $SU(2)_L\times U(1)$ gauge theory, it works very well,
but why had it to be this way? Why has the weak interaction the $V-A$ form? Why gauge symmetry and the
scalar fields at all? Renormalizability in the traditional sense is no longer a fundamental requirement and
the problem of perturbative unitarity in models with 'hand-written'
masses of vector bosons does not seem to be a very convincing argument either.
The same problem of perturbative unitarity arises in any EFT when one goes to sufficiently high energies.
On the other hand the failure of the perturbation theory does not necessarily mean that
the theory is inconsistent.
	It is puzzling, why should one generalize the gauge
symmetry principle, which turned out to be the consequence of
massless intermediate particles, to weak interaction which is
mediated by massive particles.
It would be natural to describe the massive
spin one particles of the weak interaction by an EFT without imposing gauge invariance.
More precisely, following the modern approach to QFT, i.e. the logic of EFT, what kind of theory would we construct if there did
not already exist a theory of electro-weak interactions?
Does there exist any self-consistent EFT of massive vector bosons
which is not based on the spontaneously broken (local) gauge symmetry and, if it does, how is it related to the WS model?

	The purpose of this work is to give arguments supporting the point of view that the WS
model is the way it is because any self-consistent
(parity non-conserving) leading order EFT Lagrangian of massive vector bosons interacting with fermions and electromagnetic field looks like as if it was an
$SU(2)_L\times U(1)$ gauge invariant theory with spontaneous symmetry breaking in unitary gauge with the scalar field replaced by its vacuum expectation value.

\medskip

	In trying to probe these issues, we start with analyzing the most general Lorentz invariant effective Lagrangian of massive self-interacting vector bosons. The performed analysis is similar to that of Ref.~\cite{Djukanovic:2010tb} but we do not assume the parity conservation in current work.
The most general Lorentz-invariant
effective Lagrangian contains an infinite number of interaction
terms. It is assumed that all coupling constants of
''non-renormalizable'' interactions, i.e. terms with couplings of
negative mass-dimensions, are suppressed by powers of some large
scale. Massive vector bosons are spin one particles and therefore they are described by Lagrangians with
constraints. To have a system with the right number of degrees
of freedom the coupling constants of the Lagrangian have to satisfy some non-trivial relations.
Furthermore, demanding the perturbative renormalizability in the sense of
EFT additional consistency conditions are imposed.
 In our analysis we demand the perturbative renormalizability in the sense of
EFT because if we did not include all possible couplings which absorb the
ultraviolet divergences, such interaction terms would be effectively generated by quantum corrections anyway.
On the other hand, although all loop diagrams can be made finite in any quantum field theory if we include an infinite number of counter terms in the Lagrangian, it is by no means guaranteed that these counter terms are consistent with constraints of the theory of spin one particles.
	Gauge invariant theory with the spontaneous symmetry breaking has been derived by demanding
tree-order unitarity of the $S$ matrix long time ago \cite{LlewellynSmith:1973ey,Cornwall:1973tb,Cornwall:1974km,Joglekar:1973hh}. The results of the current work are obtained by demanding perturbative renormalizability in the sense of EFT. Note that these two conditions are not equivalent. While the tree-order unitarity implies renormalizability, perturbative renormalizability in the sense of EFT is much weaker condition and it does not imply the tree-order unitarity.

First we consider an EFT of two charged massive vector bosons and show that all self-interaction terms with dimensionless coupling constants vanish. It follows from this result that a self-consistent theory of charged massive vector bosons interacting with electromagnetic field does not exist. Next we add the third, neutral vector boson and analyze such an EFT.
As a result we obtain an effective Lagrangian which is $SU(2)$ {\it locally} invariant up
to the globally invariant mass term.
	Introducing a massless fermion doublet in the above EFT and demanding
the perturbative renormalizability in the sense of EFT we obtain the leading order effective Lagrangian which is either $SU(2)$ (i.e. vector bosons interacting with vector currents) or $SU(2)_L$ (i.e. vector bosons interacting with vector minus axial vector currents) locally invariant up to the globally invariant mass term of  vector bosons. Parity non-conservation suggests that for the weak interaction the $SU(2)_L$ locally invariant Lagrangian should be chosen. As a self-consistent UV completion of the obtained leading order effective Lagrangian an $SU(2)_L$ locally invariant EFT, generalized for more fermion doublets,
is suggested.
The fermion masses and mixing are introduced by considering them as external fields. By including in the effective Lagrangian all terms which are invariant under local transformations, when the external fields are also transformed, and after choosing the external fields to be equal to the constant fermion mass matrix we obtain an perturbatively renormalizable EFT. Renormalizability is studied by analyzing the symmetries of the effective action. As the photons are  massless,  effective Lagrangian taking into account the electromagnetic interaction has to be $U(1)$ gauge invariant. Including the electromagnetic interaction in the effective Lagrangian \cite{Djukanovic:2005ag} we obtain an EFT which at the leading order coincides with the WS model in unitary gauge where the scalar field is put equal to its (constant) vacuum expectation value.

The paper is organized as follows: In section II an EFT of two charged massive vector bosons is considered. Section III deals with charged and neutral massive vector bosons. In section IV an EFT including interaction with massless fermions is analyzed. In section V the symmetries of the effective action and the perturbative renormalizability in the sense of EFT are discussed. Section VI contains the summary and general considerations.

\section{EFT of charged vector bosons}

\subsection{Lagrangian}

   Let us start with a system of two self-interacting charged massive vector bosons $V^\pm_\mu=(V^1_\mu\mp iV^2_\mu)/\sqrt{2}$ described by the most general
effective Lagrangian respecting  Lorentz invariance and charge conservation.
The effective Lagrangian contains an infinite number of terms. We assume that the coupling constants with different mass dimensions are not correlated and couplings with negative mass dimensions are suppressed by some
scale which is much larger than the energies in which we are interested. Below we treat only interaction terms
with dimensionless coupling constants starting with the Lagrangian
\begin{equation}
{\cal L}_A= {\cal L}_2+{\cal L}_4, \label{LagGeneralcv}
\end{equation}
where ${\cal L}_2$ is the free part and ${\cal L}_4$ contains all interaction terms with dimensionless coupling constants. The free part is given by
\begin{equation}
{\cal L}_2  = -{1\over 4} \ V^a_{\mu\nu} V^{a \mu\nu}
+\frac{M^2}{2} V_\mu^a V^{a \mu}, \label{kvLagrcv}
\end{equation}
where $V^a_{\mu\nu}=\partial_\mu V^a_\nu -
\partial_\nu V^a_\mu $, $M$ is the mass of vector bosons
and the summation over $a$ from 1 to 2 is implied.
   The interaction terms with dimensionless coupling constants involve four vector fields and have the form
\begin{equation}
{\cal L}_4 = - h^{a b c d} V_\mu^a V_\nu^b V^{c \mu}V^{d\nu}
,
\label{4VLcv}
\end{equation}
where $h^{a b c d}$ ($a,b,c,d=1,2$) are real parameters. Taking into account the
charge conservation these coupling constants can be written as
\begin{eqnarray} h^{1111} & = & h^{2222}=\frac{d_1+d_2}{4}\,, \nonumber\\
h^{1112} & = & -h^{1121}-h^{1211}-h^{2111},\nonumber\\
h^{1122} & = & d_2-h^{2112}-h^{1221}-h^{2211},\nonumber\\
h^{1212} & = & \frac{1}{2}\left(d_1-d_2-2\,h^{2121}\right),\nonumber\\
h^{2122} & = & -h^{1222}-h^{2212}-h^{2221} . \label{4couplingsNEWcv}
\end{eqnarray}
where $h^{1121}$,  $h^{1211}$, $h^{1221}$, $h^{1222}$,  $h^{2111}$, $h^{2112}$, $h^{2121}$,  $h^{2211}$, $h^{2212}$, $h^{2221}$  are free parameters and all other coupling constants vanish. The Lagrangian depends only on $d_1$ and $d_2$.

\subsection{Quantization}
   To quantize the above theory of massive
vector fields we use the canonical formalism following
Ref.~\cite{gitman}.
   The canonical momenta conjugated to the fields $V^a_0$ and $V^a_i$ are defined as
\begin{eqnarray}
\pi^a_0 &=& {\partial {\cal L}_A\over \partial\dot V^a_0}= 0\, \label{pi0cv},
\\
\pi^a_i &=& {\partial {\cal L}_A\over \partial\dot V^a_i}=V^a_{0i}. \label{piicv}
\end{eqnarray}
   The velocities $\dot V^a_0 $ cannot be solved from Eq.\
(\ref{pi0cv}), i.e. 
we obtain the primary constraints
\begin{equation}
  \phi_1^a=\pi^a_0. \label{phi1cv}
\end{equation}
   On the other hand, from Eq.~(\ref{piicv}) we solve
\begin{equation}
\dot V_i^a=\pi_i^a+\partial_iV_0^a. \label{aidotcv}
\end{equation}
   Next we construct the so-called total Hamiltonian density:
\begin{equation}
{\cal H}_1= \phi^a_1 z^a +{\cal H}\,, \label{hamdencv}
\end{equation} where
\begin{eqnarray}
{\cal H} & = & {\pi^{a}_i\pi^{a}_i\over 2} +\pi_i^a
\partial_i V^a_0 + \frac{1}{4} V^a_{ij}
V^a_{ij} - {M^2\over 2}V^a_\mu V^{a \mu}
+ h^{abcd} V_\mu^a V_\nu^b V^{c \mu} V^{d \nu}
\,.
\label{hamiltonian}
\end{eqnarray}
In Eq.\ (\ref{hamdencv}) $z^a$ are arbitrary functions which have to
be determined.

The primary constraints must be conserved in time. Therefore
we calculate the Poisson brackets of $\phi^a_1$ with the
Hamiltonian
 \be H_1=\int d^3{\bf x}\, {\cal
H}_1({\bf x}), \label{truehamiltoniancv} \ee and obtain
\begin{equation}
\left\{ \phi^a_1,H_1\right\}=
\partial_i\pi_i^a + M^2 V^a_0
- \left( h^{abcd}+ h^{badc}+ h^{cbad}+ h^{dcba}\right)
V_\mu^b V^c_0 V^{d \mu}
\equiv \phi_2^a=0,\ a=1,2. \label{equivphi2cv}
\end{equation}
None of the $z^b$ can be solved from
Eq.~(\ref{equivphi2cv}) and therefore $\phi_2^a$
are the secondary constraints. They also must be conserved in
time and to obtain the right number of degrees of freedom for
massive vector bosons, the $z^a$ have to be solvable from this
condition. If this is the case then no more constraints occur and the
Lagrangian describes the system with constraints of the second
class.

   Demanding the conservation of $\phi^a_2$ in time
we obtain two linear equations for the $z^a$,
\begin{equation}
 \left\{ \phi^a_2,H_1\right\} = {\cal M}^{ab} z^b +Y^a \approx
0,\quad a=1,2,
\label{zaequationscv}
\end{equation}
where the $2\times 2$ matrix ${\cal M} $ is given by
\begin{eqnarray}
{\cal M}^{ab} & = & M^2\delta^{ab}
+ \left( h^{acbd}+ h^{cadb}+ h^{bcad}+ h^{dbca}\right) V_i^c V_i^d
-\bigl(h^{abcd} +h^{badc} + h^{cbad}+h^{dcba}\nonumber\\
&+& h^{adcb}
+h^{dabc}+h^{cdab} +h^{bcda} + h^{acbd} +h^{cadb}
 +  h^{bcad} +h^{dbca}\bigr) V_0^c V_0^d,
\label{commutmatrcv}
\end{eqnarray}
and the $Y^a$ are some functions of the fields and conjugated
momenta, the particular form of which is not important in the following
discussion. If the determinant of ${\cal M}$ vanishes for some
values of the fields, then the $z^a$ cannot be determined and
additional constraints have to be imposed \cite{gitman}. This would
correspond to a wrong number of degrees of freedom. This problem, in
its various appearances, is known as the Johnson-Sudarshan
\cite{Johnson:1960vt} and the Velo-Zwanziger \cite{Velo:1969bt}
problem. To obtain a self-consistent field theory we demand
that ${\rm det} {\cal M}$ does not vanish. {  Below we analyze the
necessary conditions for the non-vanishing of the $\det \mathcal{M}$. To simplify the calculations
we calculate $\det \mathcal{M}$ for some fixed field
configurations.

For the field configurations satisfying the conditions
$$
V_i^a V_i^b=V_0^2 = 0
$$
the determinant reads
\begin{equation}
\left[(d_1+d_2) (V_0^1)^2-M^2\right] \left[3 (d_1+d_2)
(V_0^1)^2-M^2\right].
\nonumber
\end{equation}
Demanding the non-vanishing of the above expression for arbitrary
$V_0^1$ we obtain
\begin{equation}
d_1+d_2\leq0\,. \label{ineq2cv}
\end{equation}
For the field configurations satisfying the conditions
\begin{equation}
V_i^2 V_i^2 =V_i^1 V_i^2 =V_0^a= 0
\nonumber
\end{equation}
the determinant has the form
\begin{equation}
\left[M^2+(d_1+d_2) V_i^1 V_i^1\right]
\left[M^2+\left(d_1-d_2\right) \,V_i^1 V_i^1\right].
\nonumber
\end{equation}
Demanding the non-vanishing of the above expression for arbitrary
$V_i^1 V_i^1$ we obtain
\begin{equation}
d_1+d_2\geq0\,, \ d_1-d_2 \geq 0. \label{ineq4cv}
\end{equation}
From Eqs.~(\ref{ineq2cv}) and (\ref{ineq4cv}) follows that
\begin{equation}
d_2=-d_1\,,\ d_1\geq 0\,. \label{d21cv}
\end{equation}
Taking into account Eq.~(\ref{d21cv}) the determinant reads
\begin{equation}
M^4+2 \,d_1 M^2
   \left(V^1_iV^1_i+V^2_iV^2_i\right)-4 \,d_1^2 \left[(V^1_iV^2_i)^2-V^1_i V^1_i V^2_iV^2_i\right].
\end{equation}
This expression does not vanish for any field configurations for non-negative $d_1$.

    We proceed with the quantization and obtain after somewhat involved but rather
straightforward calculations the following generating functional
\begin{equation}
Z[\{J^{a\mu}\}] = \int {\cal D} V\,{\cal D}\,c\,{\cal D}\,\bar c\,{\cal D}
\lambda\,e^{i \int d^4 x \,\left( {\cal L}_{\rm
eff}+J^{a\mu}V_\mu^a\right)},
\label{GFPEffectiveCanonicalcoordinatescv}
\end{equation}
where
\begin{eqnarray}
{\cal L}_{\rm eff} & = &
{\cal L}_A - {M^2\over 2}\lambda^a \lambda^{a}
-\frac{1}{2}\,\bigl(h^{abcd}+ h^{badc}+ h^{cbad}+h^{dcba}\bigr)
\lambda^a V_i^b \lambda^{c} V_i^{d}\nonumber\\
&+& \frac{1}{2}\, \bigl(h^{abcd} +h^{badc} + h^{cbad}+h^{dcba} +h^{adcb}
+h^{dabc}+h^{cdab} +h^{bcda}\nonumber\\
& + & h^{acbd} +h^{cadb} 
+h^{bcad} +h^{dbca}\bigr) V_0^c
V_0^d\lambda^a\lambda^b \nonumber\\
&-& h^{abcd}
\left(8\,V_0^a\lambda^b\lambda^c\lambda^d
- 3\,\lambda^a\lambda^b\lambda^c\lambda^d\right)
+ M^2\,\bar c^a c^a + 4 h^{bdac}\,V_i^c
V_i^d \,\bar c^a c^b\nonumber\\
&-&  \bigl(h^{abcd} +h^{badc} + h^{cbad}+h^{dcba} +h^{adcb}
+h^{dabc}+h^{cdab} +h^{bcda} + h^{acbd} +h^{cadb} \nonumber\\& + &
h^{bcad} +h^{dbca}\bigr) \bigl(V_0^c V_0^d-\lambda^c V_0^d-V_0^c \lambda^d +\lambda^c \lambda^d\bigr)\,
\bar c^b c^a
\label{effectivelagrangiancv}
\end{eqnarray}
and the Lagrangian ${\cal L}_A$ is given by
Eqs.~(\ref{LagGeneralcv}), (\ref{kvLagrcv}), and (\ref{4VLcv}).

\subsection{Perturbative renormalizability}

	To renormalize an EFT to all orders in loop expansion the effective Lagrangian can be considered as
a Taylor series expansion in derivatives acting on fields. Divergences are absorbed order-by-order in this expansion. In the Language of Feynman diagrams this means that the vertex functions are expanded in powers of momenta and the divergences are absorbed in fields and parameters of the Lagrangian order-by-order in this expansion.

Let us analyze (some of) the necessary conditions of perturbative
renormalizability of the considered EFT of self-interacting massive vector bosons.
  We use the dimensional regularization and perform calculations including one-loop order. The
dimensional regularization puts all power-law divergences equal to zero and
parameterizes all logarithmic divergences. As the $\lambda^a$, $c^a$
and $\bar c^a$ fields do not have kinetic parts in
Eq.~(\ref{effectivelagrangian}), their contributions vanish in
perturbative calculations when the dimensional regularization is applied.

    As there is no tree order contribution in the vertex
function $V^1V^1V^1V^1$, we have to demand that the divergent part
of the corresponding one-loop contribution vanishes. Calculating the one-loop diagrams contributing in $V^1V^1V^1V^1$ vertex function we obtain the condition
\begin{eqnarray}
 d_1^2 =0.
\label{eqx}
\end{eqnarray}
We conclude that in a self-consistent EFT of two self-interacting charged vector bosons all interaction terms with dimensionless coupling constants vanish.
It follows from the last result, that a self-consistent EFT of charged vector bosons interacting with the electromagnetic field does not exist. Indeed, including the electromagnetic interaction in standard way and calculating the one-loop corrections (with virtual photons) to the $V^+V^-V^+V^-$ vertex function we obtain a  divergent expression. As the corresponding tree order diagram does not exist, one cannot get rid off this divergence.

\section{EFT of charged and neutral vector bosons}

\subsection{Lagrangian}

   Next we consider a system of three self-interacting massive vector bosons described by the most general
effective Lagrangian respecting  Lorentz invariance and charge conservation.
Two vector fields correspond to a pair of
charged particles, $V^\pm_\mu=(V^1_\mu\mp iV^2_\mu)/\sqrt{2}$, and
the third component, $V^3_\mu$, is neutral. We again assume that the coupling constants with different mass dimensions are not correlated and couplings with negative mass dimensions are suppressed by some
large scale. Here we treat only interaction terms
with dimensionless coupling constants. The considered effective Lagrangian can be written as
\begin{equation}
{\cal L}_B= {\cal L}_2+{\cal L}_3+{\cal L}_4, \label{LagGeneral}
\end{equation}
where ${\cal L}_2$ is the free Lagrangian, ${\cal L}_3$ and ${\cal L}_4$ contain interaction terms with three and four vector bosons, respectively.
The free Lagrangian is given by
\begin{equation}
{\cal L}_2  = -{1\over 4} \ V^a_{\mu\nu} V^{a \mu\nu}
+\frac{M_a^2}{2} V_\mu^a V^{a \mu} \label{kvLagr}
\end{equation}
where $V^a_{\mu\nu}=\partial_\mu V^a_\nu -
\partial_\nu V^a_\mu $, $M_a$ is the mass of $a$-th vector field
($M_1=M_2=M$),
and the summation over $a$ from 1 to 3 is implied.
   The interaction Lagrangian with three vector fields
is of the form
\begin{equation}
{\cal L}_3 = - g_V^{abc} V^a_\mu V^b_\nu \partial^\mu V^{c \nu}
- g_A^{abc}\, \epsilon^{\mu\nu\alpha\beta}  V^a_\mu V^b_\nu \partial_\alpha V^c_\beta\,,
\label{3VintLagr}
\end{equation}
where $g_V^{abc}$ and $g_A^{abc}$ ($a,b,c=1,2,3$) are coupling constants.

Using the charge conservation the coupling constants can be expressed in terms of
ten real parameters,
\begin{eqnarray}
g_V^{333} & = & g_1, \ \ \ g_V^{113}= g_2, \ \ \ g_V^{123}= -g_3, \ \
\ g_V^{213}= g_3,\nonumber\\
g_V^{223} & = & g_2, \ \ \ g_V^{311}= g_4, \ \ \ g_V^{321}= -g_5, \ \ \
g_V^{312}= g_5, \nonumber\\
g_V^{322} & = & g_4, \ \ \ g_V^{131}= g_6, \ \ \ g_V^{231}= -g_7, \ \ \
g_V^{132}= g_7,\nonumber\\ g_V^{232} & = & g_6\,,\nonumber\\
g_A^{213} & = & -g_A^{123}= g_{A1},\nonumber\\
g_A^{311} & = & g_A^{322}= -g_A^{131}= -g_A^{232}= g_{A2}, \nonumber\\
g_A^{312} & = & -g_A^{321}= -g_A^{132}= g_A^{231}= g_{A3} \,.
\label{3couplings}
\end{eqnarray}
All other constants vanish.

   The interaction Lagrangian involving four vector fields has the form
\begin{equation}
{\cal L}_4 = - h^{a b c d} V_\mu^a V_\nu^b V^{c \mu}V^{d\nu}
,
\label{4VL}
\end{equation}
where $h^{a b c d}$ ($a,b,c,d=1,2,3$) are real coupling constants
which, using the charge conservation, can be written as
\begin{eqnarray} h^{1111} & = & h^{2222}=\frac{d_1+d_2}{4}\,, \nonumber\\
h^{1112} & = & -h^{1121}-h^{1211}-h^{2111},\nonumber\\
h^{1122} & = & d_2-h^{2112}-h^{1221}-h^{2211},\nonumber\\
h^{1212} & = & \frac{1}{2}\left(d_1-d_2-2\,h^{2121}\right),\nonumber\\
h^{1323} & = & -h^{2313}-h^{3132}-h^{3231},\nonumber\\
h^{2122} & = & -h^{1222}-h^{2212}-h^{2221},\nonumber\\
h^{2323} & = & \frac{1}{2}\left(d_4-2\,h^{3232}\right),\nonumber\\
h^{3113} & = & \frac{1}{2}\left[d_3-2\,\left(h^{1133}+h^{1331}+h^{3311}\right)\right],\nonumber\\
h^{3223} & = & \frac{1}{2}\left[d_3-2\,\left(h^{2233}+h^{2332}+h^{3322}\right)\right],\nonumber\\
h^{3123} & = & -h^{1233}-h^{1332}-h^{2133}-h^{2331} -  h^{3213}-h^{3312}-h^{3321},\nonumber\\
h^{3131} & = & \frac{1}{2}\left(d_4-2\,h^{1313}\right),\nonumber\\
h^{3333} & = & d_5\,. \label{4couplingsNEW}
\end{eqnarray}
where $h^{1121}$, $h^{1133}$, $h^{1211}$, $h^{1221}$, $h^{1222}$, $h^{1233}$, $h^{1313}$, $h^{1331}$, $h^{1332}$, $h^{2111}$, $h^{2112}$, $h^{2121}$, $h^{2133}$, $h^{2211}$, $h^{2212}$, $h^{2221}$, $h^{2233}$, $h^{2313}$, $h^{2331}$, $h^{2332}$, $h^{3132}$, $h^{3231}$, $h^{3213}$, $h^{3232}$, $h^{3311}$, $h^{3312}$, $h^{3321}$, $h^{3322}$ are free parameters and all other coupling constants vanish. The Lagrangian depends only on $d_1,\cdots,d_5$.

\subsection{Quantization}
   As above, to quantize the considered theory we use the canonical formalism following
Ref.~\cite{gitman}. The analysis below is similar to the one of Ref.~\cite{Djukanovic:2010tb}, with the difference that in Ref.~\cite{Djukanovic:2010tb}
the parity conservation has been taken as an input.
   The canonical momenta conjugated to the fields $V^a_0$ and $V^a_i$ are defined as
\begin{eqnarray}
\pi^a_0 &=& {\partial {\cal L}_B\over \partial\dot V^a_0}=-g_V^{bca}
V^b_0 V^c_0\, \label{pi0},
\\
\pi^a_i &=& {\partial {\cal L}_B\over \partial\dot V^a_i}=V^a_{0i}
+g_V^{bca} V^b_0 V^c_i+g_A^{bca}\epsilon^{ijk0} V^b_j V^c_k\,. \label{pii}
\end{eqnarray}
   The velocities $\dot V^a_0 $ cannot be solved from Eq.\
(\ref{pi0}), i.e. 
we obtain the primary constraints
\begin{equation}
  \phi_1^a=\pi^a_0+g_V^{bca} V^b_0 V^c_0. \label{phi1}
\end{equation}
   On the other hand, from Eq.~(\ref{pii}) we solve
\begin{equation}
\dot V_i^a=\pi_i^a+\partial_iV_0^a-g_V^{bca}V_0^b V_i^c
- g_A^{bca}\epsilon^{ijk0} V^b_j V^c_k. \label{aidot}
\end{equation}
   Next we construct the so-called total Hamiltonian density:
\begin{equation}
{\cal H}_1= \phi^a_1 z^a +{\cal H}\,, \label{hamden}
\end{equation} where
\begin{eqnarray}
{\cal H} & = & {\pi^{a}_i\pi^{a}_i\over 2} +\pi_i^a
\partial_i V^a_0 + \frac{1}{4} V^a_{ij}
V^a_{ij} - {M_a^2\over 2}V^a_\mu V^{a \mu}
-g_V^{abc}V^a_0 V^b_i \pi^c_i - g_A^{abc}\epsilon^{ijk0}V^a_j V^b_k \pi^c_i\nonumber\\
&-& g_V^{abc} V_0^a V_i^b
\partial_iV^c_0-g_V^{abc} V_i^a V_0^b
\partial_iV^c_0
+ g_V^{abc} V_i^a V_j^b \partial_i V^c_j+\frac{1}{2} g_V^{abc}
g_V^{a'b'c} V_0^a V_i^b
V^{a'}_0 V_i^{b'} \nonumber\\
&+& \frac{1}{2} g_A^{bca}
g_A^{b'c'a} \epsilon^{ijk0}\epsilon^{i j' k' 0}\,V_j^b V_{j'}^{b'} V^{c}_k V_{k'}^{c'}
+ g_A^{abc}
g_V^{b'c'c} \epsilon^{ijk0}\,V_j^a V_{k}^b V^{b'}_0 V_{i}^{c'}\nonumber\\
&+&g_A^{abc} \epsilon^{ijk0} V^a_0 V^b_j \partial_i V_k ^c-g_A^{abc} \epsilon^{ijk0} V^a_j V^b_0
\partial_i V_k ^c
+ h^{abcd} V_\mu^a V_\nu^b V^{c \mu} V^{d \nu}
\,.
\label{hamiltonian}
\end{eqnarray}
In Eq.\ (\ref{hamden}) $z^a$ are arbitrary functions which have to
be determined.

To demand the conservation of the primary constraints in time 
we calculate the Poisson brackets of $\phi^a_1$ with the
Hamiltonian
 \be H_1=\int d^3{\bf x}\, {\cal
H}_1({\bf x})\,, \label{truehamiltonian} \ee and obtain
\begin{eqnarray}
\left\{ \phi^a_1,H_1\right\}&=&
\left(g_V^{bca}+g_V^{cba}-g_V^{acb}-g_V^{cab}\right) V_0^c z^b
+ \partial_i\pi_i^a+g_V^{abc}V_i^b\pi_i^c+
\left( g_V^{abc}+g_V^{bac}\right)\,V^b_i\partial_iV^c_0\nonumber\\
& - & g_V^{bca}\partial_i\left( V^b_0V^c_i\right) -
g_V^{cba}\partial_i\left( V^b_0V^c_i\right)+ M_{a}^2 V^a_0 
- g_V^{abc} g_V^{a'b'c} V_i^b V^{a'}_0 V_i^{b'} \nonumber\\ &-&
g_A^{a'bc} g_V^{ac'c} \epsilon^{ijk0}\,V_j^{a'} V_{k}^b V_{i}^{c'}
-g_A^{abc} \epsilon^{ijk0} V^b_j \partial_i V_k ^c+g_A^{b a c} \epsilon^{ijk0} V^{b}_j
\partial_i V_k ^c
\nonumber\\
&-& \left( h^{abcd}+ h^{badc}+ h^{cbad}+ h^{dcba}\right)
V_\mu^b V^c_0 V^{d \mu}
\equiv A^{ab}z^b+\chi^a,\quad a=1,2,3. \label{equivphi2}
\end{eqnarray}
Using Eq.\ (\ref{3couplings}), defining $\gamma_1=g_5+g_7$ and
$\gamma_2=g_4+g_6-2g_2$, the matrix $A$ is given by
\begin{equation}
\label{defA} A=\left(\begin{array}{ccc} 0&-2\gamma_1 V_0^3&
\gamma_2 V_0^1-\gamma_1 V_0^2\\
2\gamma_1V_0^3 & 0 & \gamma_1 V_0^1+\gamma_2 V_0^2\\
-(\gamma_2 V_0^1-\gamma_1 V_0^2)& -(\gamma_1 V_0^1+\gamma_2 V_0^2)&0
\end{array}\right).
\end{equation}
The determinant of $A$ vanishes and therefore the system of equations
\begin{equation}
A^{a b} z^b= -\chi^a \label{equations1}
\end{equation}
can be satisfied only if the right-hand sides satisfy the secondary constraint
\begin{equation}
\phi_2= \chi^1\,(\gamma_1 V_0^1+\gamma_2 V_0^2) + \chi^2\, (\gamma_1
V_0^2-\gamma_2 V_0^1)-\chi^3\, 2 \gamma_1\, V_0^3 = 0\,.
\label{additionalconstraint1}
\end{equation}
Let us consider Eq.~(\ref{equations1}) for the case where at least
one of $\gamma_1$ or $\gamma_2$ does not vanish. For non-vanishing
$V_0^1$ and/or $V_0^2$ we obtain
\begin{eqnarray}
z^1 & = & \frac{\chi_3+\gamma_1 z^2\,V_0^1+\gamma_2 \,z^2
V_0^2}{\gamma_1\, V_0^2-\gamma_2\, V_0^1},\nonumber\\
z^3 & = & \frac{\chi_1+2\,
\gamma_1\,z^2\,V_0^3}{\gamma_2\,V_0^1-\gamma_1\,V_0^2}
\label{zsolutions}
\end{eqnarray}
and $z^2$ can be solved from the time conservation of the constraint of
$\phi_2$, $\left\{ \phi_2,H_1\right\}= 0$.
   However, in this case we obtain a wrong number of constraints
of the second class \cite{second_class} for our system of three
massive vector fields - four instead of six.
   Therefore for a self-consistent theory we have to require
\begin{equation}
g_7= - g_5\,, \ \ \  2g_2=g_4+g_6\,. \label{gfirstconstraint}
\end{equation}
   In this case none of the $z^b$ can be solved from
Eq.~(\ref{equivphi2}) and
\begin{eqnarray}
\left\{ \phi^a_1,H_1\right\}&=&
\partial_i\pi_i^a+g_V^{abc}V_i^b\pi_i^c+
\left( g_V^{abc}+g_V^{bac}\right)\,V^b_i\partial_iV^c_0
-g_V^{bca}\partial_i\left( V^b_0V^c_i\right) -
g_V^{cba}\partial_i\left( V^b_0V^c_i\right)+ M_{a}^2 V^a_0 \nonumber\\ &-&
 g_V^{abc} g_V^{a'b'c} V_i^b V^{a'}_0 V_i^{b'} 
- g_A^{a'bc}
g_V^{ac'c} \epsilon^{ijk0}\,V_j^{a'} V_{k}^b V_{i}^{c'}
-g_A^{abc} \epsilon^{ijk0} V^b_j \partial_i V_k ^c+g_A^{b a c} \epsilon^{ijk0} V^{b}_j
\partial_i V_k ^c
\nonumber\\
&-&   \left( h^{abcd}+ h^{badc}+ h^{cbad}+ h^{dcba}\right)
V_\mu^b V^c_0 V^{d \mu}
\equiv \phi_2^a,\quad a=1,2,3,
\label{equivphi22}
\end{eqnarray}
are the secondary constraints. They also must be conserved in
time and  the $z^a$ have to be solvable from this
condition. If this is the case then no more constraints occur and the
Lagrangian describes the system with the right number of constraints of the second
class.

   Demanding the conservation of $\phi^a_2$ in time and taking Eq.~(\ref{gfirstconstraint}) into account,
we obtain a system of three linear equations for the $z^a$,
\begin{equation}
 \left\{ \phi^a_2,H_1\right\} = {\cal M}^{ab} z^b +Y^a =
0,\quad a=1,2,3,
\label{zaequations}
\end{equation}
where the $3\times 3$ matrix ${\cal M} $ is given by
\begin{eqnarray}
{\cal M}^{ab} & = & M^2_a\delta^{ab}
-\left(g_V^{bca}+g_V^{cba}\right)\partial_i V_i^c - \bigl[ g_V^{ace} g_V^{bde} 
- \left( h^{acbd}+ h^{cadb}+ h^{bcad}+ h^{dbca}\right) \bigr] V_i^c V_i^d\nonumber\\
&-& \bigl(h^{abcd} +h^{badc} + h^{cbad}+h^{dcba} +h^{adcb}
+h^{dabc}+h^{cdab} +h^{bcda} + h^{acbd} +h^{cadb} \nonumber\\
& + & h^{bcad} +h^{dbca}\bigr) V_0^c V_0^d,
\label{commutmatr}
\end{eqnarray}
and the  particular form of $Y^a$ is not important in the following
discussion. As in the above case of two charged vector bosons, to obtain a self-consistent field theory we demand
that ${\rm det} {\cal M}$ does not vanish. Again, we calculate $\det \mathcal{M}$ only for some fixed field
configurations starting with the configurations satisfying the conditions
$$
V_i^a V_i^b = V_0^a=
\partial_i V_i^1= \partial_i V_i^2= 0.
$$
The corresponding determinant reads
$$
\left(M^2-2 \,g_2 \,\partial_i V_i^3\right)^2 \left(M_3^2-2 \,g_1
\,\partial_i V_i^3\right).
$$
Demanding the non-vanishing of the above expression for arbitrary
$\partial_i V_i^3$ we obtain
\begin{equation}
g_1=g_2=0\,. \label{g1g2}
\end{equation}
Next we consider the field configurations satisfying the conditions
$$
V_i^a V_i^b=V_0^2= V_0^1= 0
$$
and obtain the determinant
$$
\left[M^2-(d_3+d_4) \,(V_0^3)^2\right]^2
   \left[M_3^2-12\, d_5 \,(V_0^3)^2\right].
$$
Demanding the non-vanishing of the above expression for arbitrary
$V_0^3$ we obtain
\begin{equation}
d_3+d_4\leq0\,,\ \ d_5\leq0\,. \label{ineq1}
\end{equation}
For the field configurations satisfying the conditions
$$
V_i^a V_i^b=V_0^2=V_0^3 = 0
$$
the determinant reads
\begin{equation}
\left[(d_1+d_2) (V_0^1)^2-M^2\right] \left[3 (d_1+d_2)
(V_0^1)^2-M^2\right]
\left[M_3^2-(d_3+d_4) (V_0^1)^2\right].
\nonumber
\end{equation}
Demanding the non-vanishing of the above expression for arbitrary
$V_0^1$ we obtain
\begin{equation}
d_1+d_2\leq0\,. \label{ineq2}
\end{equation}
Further we take the field configurations satisfying the conditions
\begin{equation}
V_i^1 V_i^1 = V_i^1V_i^2=V_i^1V_i^3=V_i^2 V_i^2
= V_i^1 V_i^2=
V_i^2 V_i^3= V_0^a=0
\end{equation}
and obtain for the determinant
$$
\left(M_3^2+4\, d_5 \,V_i^3 V_i^3\right)
   \left[M^2+\left(d_4 -g_4^2-g_5^2\right)
   \,V_i^3 V_i^3\right]^2.
$$
Demanding the non-vanishing of the above expression for arbitrary
$V_i^3V_i^3$ we obtain
\begin{equation}
d_5\geq0\,, \ d_4 \geq g_4^2+g_5^2. \label{ineq3}
\end{equation}
From Eqs.~(\ref{ineq1}) and (\ref{ineq3}) follows that
\begin{equation}
d_5=0\,. \label{d52}
\end{equation}
For the field configurations satisfying the conditions
\begin{equation}
V_i^1 V_i^3=V_i^2 V_i^3=V_i^3 V_i^3= V_i^2 V_i^2
=V_i^2 V_i^3=V_i^1 V_i^2 =V_0^a= 0
\nonumber
\end{equation}
the determinant reads
\begin{equation}
\left[M^2+(d_1+d_2) V_i^1 V_i^1\right]
\left[M^2+\left(d_1-d_2-g_3^2\right) \,V_i^1 V_i^1\right]
\left[M_3^2+\left(d_4-g_4^2-g_5^2\right) \,V_i^1 V_i^1\right].
\nonumber
\end{equation}
Demanding the non-vanishing of the above expression for arbitrary
$V_i^1 V_i^1$ we obtain
\begin{equation}
d_1+d_2\geq0\,, \ d_1-d_2 \geq g_3^2. \label{ineq4}
\end{equation}
From Eqs.~(\ref{ineq2}) and (\ref{ineq4}) follows that
\begin{equation}
d_2=-d_1\,,\ d_1\geq g_3^2/2\,. \label{d21}
\end{equation}
Next, the field configurations satisfying the conditions
$$
V_i^a V_i^b= 0,\ V_0^3= V_0^2\,, \ V_0^1= 0
$$
lead to the following expression of the determinant
\begin{equation}
\left[(d_3+d_4) \,(V_0^2)^2-M^2\right] \bigl\{\left[(d_3+d_4)
\,(V_0^2)^2-M_3^2\right] M^2
+(d_3+d_4) \,(V_0^2)^2 \left[ M_3^2+3
(d_3+d_4) \,(V_0^2)^2\right]\bigr\}.
\nonumber
\end{equation}
If $d_4\neq -d_3$ the above determinant vanishes for
$$ (V_0^2)^2=
   -\frac{M^2+M_3^2+\sqrt{M^4+14 M_3^2 M^2+M_3^4}}{6
\,(d_3+ d_4)}\,.
$$
Taking into account Eq.~(\ref{ineq1}) we conclude  that the following condition has to be satisfied
\begin{equation}
d_4=-d_3\,. \label{d42}
\end{equation}
It can be shown that no further constraints on couplings are imposed by demanding the non-vanishing of  ${\rm det} {\cal M}$
(see appendix in Ref.~\cite{Djukanovic:2010tb}).

    We proceed with the quantization and obtain the following generating functional
\begin{equation}
Z[\{J^{a\mu}\}] = \int {\cal D} V\,{\cal D}\,c\,{\cal D}\,\bar c\,{\cal D}
\lambda\,e^{i \int d^4 x \,\left( {\cal L}_{\rm
eff}+J^{a\mu}V_\mu^a\right)},
\label{GFPEffectiveCanonicalcoordinates}
\end{equation}
where
\begin{eqnarray}
{\cal L}_{\rm eff} & = &
{\cal L}_B - {M_a^2\over 2}\lambda^a \lambda^{a}- \frac{1}{2}
\left( g^{acb}+g^{cab}\right) \partial_i V_i^c\lambda^a \lambda^b
+\frac{1}{2}\,\bigl( g^{abe} g^{cde}- h^{abcd}- h^{badc}- h^{cbad}\nonumber\\
&-& h^{dcba}\bigr)
\lambda^a V_i^b \lambda^{c} V_i^{d}
+ \frac{1}{2}\, \bigl(h^{abcd} +h^{badc} + h^{cbad}+h^{dcba} +h^{adcb}
+h^{dabc}+h^{cdab} +h^{bcda}\nonumber\\
& + & h^{acbd} +h^{cadb} 
+h^{bcad} +h^{dbca}\bigr) V_0^c
V_0^d\lambda^a\lambda^b + g^{bca}\partial_0 \lambda^{a }
\lambda^b\lambda^c \nonumber\\
&-& h^{abcd}
\left(8\,V_0^a\lambda^b\lambda^c\lambda^d
- 3\,\lambda^a\lambda^b\lambda^c\lambda^d\right)
+ M^2_a\,\bar c^a c^a - \left( g^{ace} g^{bde} - 4 h^{bdac}
\right) \, V_i^c
V_i^d \,\bar c^a c^b\nonumber\\
&-&  \bigl(h^{abcd} +h^{badc} + h^{cbad}+h^{dcba} +h^{adcb}
+h^{dabc}+h^{cdab} +h^{bcda} + h^{acbd} +h^{cadb} \nonumber\\& + &
h^{bcad} +h^{dbca}\bigr) \bigl(V_0^c V_0^d-\lambda^c V_0^d-V_0^c \lambda^d +\lambda^c \lambda^d\bigr)\,
\bar c^b c^a
\label{effectivelagrangian}
\end{eqnarray}
and the Lagrangian ${\cal L}_B$ is given by
Eqs.~(\ref{LagGeneral})-(\ref{3VintLagr}), and (\ref{4VL}).

\subsection{Perturbative renormalizability}

  Below we analyze (some of) the necessary conditions of perturbative
renormalizability using the dimension regularization
and performing calculations including one-loop order. As the $\lambda^a$, $c^a$
and $\bar c^a$ fields do not have kinetic parts in
Eq.~(\ref{effectivelagrangian}), their contributions vanish in
perturbative calculations when the dimensional regularization is applied.

\begin{figure}
\epsfig{file=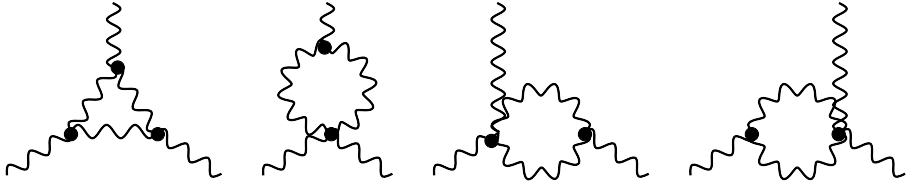, width=12truecm}
\caption[]{\label{VVV:fig} One-loop contributions to the
three-vector vertex function. The wiggly line corresponds to the
vector-meson.}
\end{figure}

We start with vertex functions of three vector-bosons  shown in
Fig.~\ref{VVV:fig}. For $V^3V^3V^3$ vertex function there is no tree order diagram, therefore
the divergent part of the sum of the corresponding one-loop diagrams has to vanish leading to the condition \cite{thanks}
\begin{equation}
g_4(3 d_3+5 g_4^2+3 g_5^2)=0.
\label{V3V3V3}
\end{equation}
Analogously to the above case we demand that the divergent part of the vertex function of four vector-bosons  $V^3V^3V^3V^3$
(shown in Fig.~\ref{VVVV:fig}) vanishes. This leads to
\begin{equation}
15 \,D_1^2+384\,D_1\,g_{A2}^2+1280 \,g_{A2}^4=0\,, \label{condition2}
\end{equation}
where
\begin{equation}
D_1=-d_3-g_5^2.
\label{D1Def}
\end{equation}
It follows from Eqs.~(\ref{ineq3}) and (\ref{d42}) that $D_1$ is non-negative. Therefore Eq.~(\ref{condition2}) leads to
\begin{eqnarray}
g_4 & = & 0,\nonumber\\
g_{A2}&=& 0,\nonumber\\
d_3 &=&-g_5^2. \label{d3g4}
\end{eqnarray}
Taking into account Eq.~(\ref{d3g4}) and demanding that the
divergent part of the one-loop contribution in $V^3V^3V^1V^1$ vertex
function  has the same tensor structure as the tree-order diagram, and hence can be absorbed in the
renormalization of the corresponding
coupling constant, we obtain
\begin{equation}
g_{5}^2 \left[2
   (\,g_3+\,g_{5})^2 M_3^2
   \,M^2+\left(\,g_3
   M_3^2+\,g_{5}
   \,M^2\right)^2+8 M^2M_3^2(g_{A1}+g_{A3})^2\right]=0\,,
\label{V3311Conditions}
\end{equation}
It follows from Eq.~(\ref{V3311Conditions}) that
\begin{eqnarray}
{\rm either} \ \ M_3 & = & M\,,\ \  g_5 = -g_3\,, \ \ {\rm and} \ \ g_{A3}=-g_{A1} \ \ {\rm or} \ \ g_5
=0. \label{eqalternative}
\end{eqnarray}

\begin{figure}
\epsfig{file=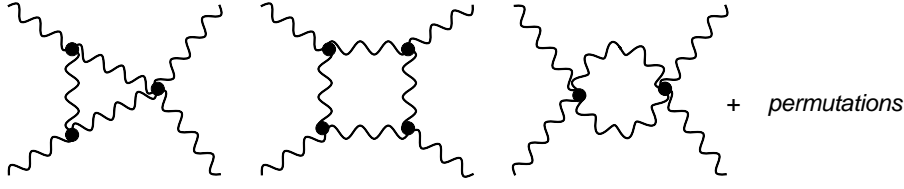, width=12truecm}
\caption[]{\label{VVVV:fig} One-loop contributions to the
four-vector vertex function. The wiggly line corresponds to the
vector-meson.}
\end{figure}

    Next, as there is no tree order contribution in the vertex
function $V^1V^1V^1V^1$, we have to demand that the divergent part
of the corresponding one-loop contribution vanishes. This gives, taking into
account the values of $d_3$ and $g_4$ from Eq.~(\ref{d3g4}),
\begin{eqnarray}
&& 60 d_1^2 M_3^4 M^4+8 g_3^3 g_5
   M_3^4 \left(M_3^2-M^2\right) M^2+g_3^4 M_3^4 \left(M^4-4 M_3^2
   M^2+15 M_3^4\right)\nonumber\\
&& -4 d_1 \biggl[M_3^4 \left(g_3^2 \left(3 M^4-2 M_3^2 M^2+5
   M_3^4\right)-48 M^4
   \left(g_{\text{A1}}+g_{\text{A3}}\right){}^2\right)+2 g_3 g_5 M^2
   \left(M_3^2-4 M^2\right) M_3^4 \nonumber\\
   && +g_5^2 \left(5 M^4 M_3^4-2 M^6
   M_3^2\right)\biggr]
   +5 M^4 \biggl[16 g_5^2 M_3^4
   \left(g_{\text{A1}}+g_{\text{A3}}\right){}^2+32 M_3^4
   \left(g_{\text{A1}}+g_{\text{A3}}\right){}^4\nonumber\\
   && +g_5^4 \left(M^4+2
   M_3^4\right)\biggr]
   +4 g_3^2 M_3^2 M^4 \biggl[g_5^2
   \left(M_3^2-M^2\right)-4 M_3^2
   \left(g_{\text{A1}}+g_{\text{A3}}\right){}^2\biggr]\nonumber\\
   && +4 g_3 g_5 M_3^2
   M^4 \biggl[8 M_3^2 \left(g_{\text{A1}}+g_{\text{A3}}\right){}^2+g_5^2
   \left(M^2+2 M_3^2\right)\biggr]=0.
\label{eqx}
\end{eqnarray}
From the first solution in Eq.~(\ref{eqalternative}) and from Eq.~(\ref{eqx}) we obtain
\begin{eqnarray}
d_1 & = & g_3^2/2,
\ \ g_5 =  -g_3, \ \ g_{A3}=-g_{A1}, \ \ M_3=M. \label{d1eq}
\end{eqnarray}
We also need to consider the second solution in Eq.~(\ref{eqalternative}), i.e. $g_5=0$. For this case we calculate the renormalization of the masses
of vector bosons and obtain that the
mass of the neutral particle does not get renormalized, and for the charged particles we obtain an infinite renormalization proportional to
\begin{equation}
6 \,d_1 M^4-g_3^2 \left(M_3^2
   M^2-M_3^4+M^4\right)+8 (g_{A1}+g_{A3})^2 M^2
   \left(M_3^2+M^2\right).
\label{mcren}
\end{equation}
If $M=\alpha M_3$, where $\alpha$ is a pure number, then the expression in Eq.~(\ref{mcren}) has to vanish because $M_3$ does not get renormalized.
However the solution for $d_1$ obtained from this condition does not satisfy
Eq.~(\ref{eqx}) for real values of $\alpha$, $g_3$ and $g_{A1}+g_{A3}$.
For the case when $\alpha$ is not a pure number we obtain from Eq.~(\ref{eqx})
\begin{eqnarray}
d_1 & = & \frac{g_3^2 \left(3
   M^4-2 M_3^2 M^2+5 M_3^4\right)-48 M^4 \left(g_{\text{A1}}+g_{\text{A3}}\right){}^2\pm\sqrt{D}}{30 M^4},\nonumber\\
   D & = & g_3^4 \left(-6 M^8+48
   M_3^2 M^6-191 M_3^4 M^4-20 M_3^6 M^2+25 M_3^8\right) \nonumber\\
   &-& 96
   M^8 \left(g_{\text{A1}}+g_{\text{A3}}\right){}^4-48 g_3^2
   \left(M^4-4 M_3^2 M^2+10 M_3^4\right) M^4
   \left(g_{\text{A1}} +g_{\text{A3}}\right){}^2,
\label{d1not}
\end{eqnarray}
which is not compatible
with the assumption that parameters with different mass dimensions are
not correlated (because it relates the masses and dimensionless
coupling constants). Moreover, Eq.~(\ref{d1not}) leads to a highly algebraic condition for the renormalization of parameters $M$, $g_3$ and $g_{A1}+g_{A3}$ which cannot be solved in perturbation theory.

\medskip

Finally, all obtained relations among coupling constants and masses can be
written as
\begin{eqnarray}
g_V^{abc} & = & -g_3 \,\epsilon^{abc} \,,\nonumber\\
g_A^{abc} & = & g_{A1} \,\epsilon^{abc} \,,\nonumber\\
h^{abcd} & = &
\frac{1}{4}\,g_V^{abe}g_V^{cde}\,,\nonumber\\
M_1 & = & M_2=M_3=M \,.\label{ccconditions1}
\end{eqnarray}


For the couplings in Eq.~(\ref{ccconditions1}) the determinant of the matrix of Poisson brackets of constraints
$ \left\{ \phi_l,\phi_{l'}\right\}$ does not depend on
fields, i.e.\ ghost fields completely decouple from vector boson
fields. The matrix ${\cal M}^{ab}$ in Eq.~(\ref{zaequations})
becomes $M^2 \delta^{ab}$, i.e. all $z^a$ are solved for {\it
all} field configurations.
   Denoting $g_3=g$ the effective Lagrangian of Eq.~(\ref{effectivelagrangian})
can be written in a compact form
\begin{eqnarray}
{\cal L}_{\rm eff} & = & {\cal L}_A=-{1\over 4} \ G^a_{\mu\nu} G^{a
\mu\nu} +\frac{M^2}{2} V_\mu^a V^{a \mu}
-g_{A 1}\epsilon^{abc}\, \epsilon^{\mu\nu\alpha\beta}  V^a_\mu
V^b_\nu
\partial_\alpha V^c_\beta\,, \label{EffLagr}
\end{eqnarray}
where
\begin{equation}
G^a_{\mu\nu}=V^a_{\mu\nu}-g\,\epsilon^{bca}V^b_\mu V^c_\nu\,.
\label{gdefinition}
\end{equation}
The last term in Eq.~(\ref{EffLagr}) can be
written as a total derivative. As the Lagrangian contains the mass
term of the vector bosons, the contributions of the slowly decaying
(instanton-like) configurations to the path integral vanish,
therefore the total derivative term can be dropped. Unlike the
massless standard Yang-Mills theory this term does not contribute in physical quantities.

The final generating functional of the Green's functions has the
form
\begin{equation}
Z[\{J^{a\mu}\}] = \int {\cal D} V\,e^{i \int d^4 x \,\left( {\cal
L}_A+J^{a\mu}V^a_\mu\right)} \label{GFFinal}
\end{equation}
and leads to ''naive'' Feynman rules.


\section{Inclusion of fermions}

In this section we include in the above considered EFT a doublet of massless fermions interacting with vector bosons.

Taking into account the results of the previous section the most general
Lagrangian of a couple of (charged) fermion fields
\begin{equation}
\label{fermiondefinition} \Psi = \left(
\begin{array}{l} \psi_u \\
\psi_d
\end{array}
\right)
\end{equation}
interacting with vector fields can be written as
\begin{equation}
{\cal L}_{\rm F}=-{1\over 4} \ G^a_{\mu\nu} G^{a
\mu\nu} +\frac{M^2}{2} V_\mu^a V^{a \mu} +i \bar\Psi
\partial\hspace{-.55em}/\hspace{.1em} \Psi
+\frac{1}{2}\,\bar\Psi \left(\gamma^\mu\,T^a_V+\gamma^5
\gamma^\mu\,T^a_A \right)\Psi\,V^a_\mu\,+\cdots, \label{EffLagrWithFermions}
\end{equation}
where we have shown explicitly only terms containing interactions with dimensionless
coupling constants with $T_A^a$ and $T_V^a$ the matrices of axial vector and vector coupling constants.
These matrices can be written in terms of real parameters as follows
\begin{eqnarray}
T_{V}^1 & = & \left(
\begin{array}{ll}
 0 & \rho_{ud}^V \\
 \rho_{ud}^V & 0
\end{array}
\right),\nonumber\\
T_V^2 & = & \left(
\begin{array}{ll}
 0 & -i \rho_{ud}^V \\
 i \rho_{ud}^V & 0
\end{array}
\right),\nonumber\\
T_V^3 & = & \left(
\begin{array}{ll}
 \rho_{3}^V+g_3^V & 0 \\
 0 & -\rho_{3}^V+g_3^V
\end{array}
\right),\nonumber\\
T_A^1 & = & \left(
\begin{array}{ll}
0 & \rho_{ud}^A \,e^{i \theta_{ud}^A}\\
\rho_{ud}^A \,e^{-i \theta_{ud}^A} & 0
\end{array}
\right),\nonumber\\
T_A^2 & = & \left(
\begin{array}{ll}
0 & -i \rho_{ud}^A \,e^{i \theta_{ud}^A}  \\
i \rho_{ud}^A \,e^{-i \theta_{ud}^A} & 0
\end{array}
\right),\nonumber\\
T_A^3 & = & \left(
\begin{array}{ll}
\rho_{3}^A+g_3^A  & 0 \\
 0 & -\rho_{3}^A+g_3^A
\end{array}
\right), \label{Tmatrices}
\end{eqnarray}
where the phase of the non-diagonal vector coupling has been excluded by
redefining the $\psi_u$ and $\psi_d$ fields.

It is straightforward to quantize the leading order Lagrangian of Eq.~(\ref{EffLagrWithFermions}) of massive
vector fields interacting with fermions using the canonical formalism of
Ref.~\cite{gitman}. The resulting generating functional of Green's functions leads to ''naive'' Feynman rules with
ghost fields decoupled from the fermion and vector boson fields.

Analogously to the previous section below we analyze the conditions of perturbative renormalizability to one-loop order.
First we calculate the divergent part of the fermion loop contribution to the vector boson self-energy and obtain
\begin{equation}
\Pi^{\mu\nu}_{\rm ab,div} = -\,\frac{ 2 \bar\lambda}{3}
 \sum_{i,j=1}^2 \Biggl[ \left( p^2 g^{\mu \nu } - p^{\mu } p^{\nu }\right)
\left(  T_{V,ji}^a\,T_{V,ij}^b  +  T_{A,ji}^a T_{A,ij}^b\right)\Biggr],
\label{VSEFloop}
\end{equation}
where $\bar\lambda\sim \frac{1}{n-4}$ parameterizes the divergence with $n$ the number of space-time dimensions.
From Eq.~(\ref{VSEFloop}) we obtain the renormalization factors for vector boson fields.

Next, calculating the fermion loop contributions to the divergent parts of vertex functions and demanding that the fermion loop
contributions in $V^1V^2V^3$ and $V^1V^2V^1V^2$ vertex functions lead to the
same renormalization of the coupling $g$ we obtain the following equation
\begin{equation}
g^2 \left[(g^A_3)^2+(g^V_3)^2\right]+\left[g \rho^A_3-4
\rho^A_{ud} \,\rho^V_{ud} \cos (\theta^A_{ud})\right]^2
+\left[g \,\rho^V_3-2 \left((\rho^A_{ud})^2+(\rho^V_{ud})^2\right)\right]^2=0\,.
\label{conCond1}
\end{equation}
From Eq.~(\ref{conCond1}) we deduce that:
\begin{eqnarray}
g_3^V & = & g_3^A=0,\nonumber\\
\rho^A_{3} & = & \frac{4\,
\rho^A_{ud} \,\rho^V_{ud} \cos (\theta^A_{ud})}{g},\nonumber\\
\rho^V_{3} & = & \frac{2 \left[(\rho^A_{ud})^2+(\rho^V_{ud})^2\right]}{g}.
\label{solvcond1}
\end{eqnarray}
Next condition is obtained by demanding that the fermion loop contributions to $V^1V^1V^3V^3$ and $V^1V^2V^1V^2$ vertex functions lead to the
same renormalization of the coupling $g$. Using Eq.~(\ref{solvcond1}) we are lead to
\begin{eqnarray}
&& 16 \bigl[(\rho^A_{ud})^6+9 (\rho^A_{ud})^4 (\rho^V_{ud})^2
+9 (\rho^A_{ud})^2 (\rho^V_{ud})^4+(\rho^V_{ud})^6\bigr]
+8 (\rho^A_{ud})^2 (\rho^V_{ud})^2 \cos (2 \,\theta^A_{ud}) \bigl[12 (\rho^A_{ud})^2\nonumber\\
&& +12 (\rho^V_{ud})^2-g^2\bigr]+ g^4 \left[(\rho^A_{ud})^2+(\rho^V_{ud})^2\right]
-8\,g^2 \left[(\rho^A_{ud})^4
+5 (\rho^A_{ud})^2 (\rho^V_{ud})^2+(\rho^V_{ud})^4\right]=0.
\label{ConsCond2}
\end{eqnarray}

Next we calculate the one-loop contributions in the fermion self-energy and in the $V \bar\Psi \Psi$ vertex function. We find that the
divergent part of the self-energy diagram, contributing in the fermion field renormalization, vanishes. Divergent part of the
vertex diagram, contributing in the renormalization of couplings in Eq.~(\ref{Tmatrices}) also vanishes. Hence the renormalization
of these coupling constants is given by renormalization factors of the vector boson fields.

    By taking into account Eq.~(\ref{solvcond1}) and writing the remaining bare parameters as
\begin{eqnarray}
g & = & g_R+\delta g,\nonumber\\
\rho^A_{ud} & = & \rho^A_{ud R}+\delta \rho^A_{ud},\nonumber\\
\rho^V_{ud} & = & \rho^V_{ud R}+\delta \rho^V_{ud},\nonumber\\
\theta^A_{ud} & = & \theta^A_{ud R}+\delta \theta^A_{ud},
\label{bare}
\end{eqnarray}
and demanding that the renormalization of vertices $\gamma^\mu\,T^1_V+\gamma^5
\gamma^\mu\,T^1_A $ and $\gamma^\mu\,T^3_V+\gamma^5
\gamma^\mu\,T^3_A $ are consistent with each other we obtain:
\begin{eqnarray}
\delta \rho^A_{ud} & = & \frac{\bar\lambda}{6}   \,\rho^A_{udR}
\left[\left(\rho_{ud R}^A\right)^2+\left(\rho_{ud R}^V\right)^2\right],\nonumber\\
\delta \rho^V_{ud} & = &  \frac{\bar\lambda}{6} \,\rho^V_{udR}
\left[\left(\rho_{ud R}^A\right)^2+\left(\rho_{ud R}^V\right)^2\right],\nonumber\\
\delta \theta^A_{ud} & = & 0.
\label{ctts}
\end{eqnarray}
In deriving Eq.~(\ref{ctts}) the value for the one-loop order counter term
\begin{eqnarray}
\delta g_1 & = & \frac{\bar\lambda }{3 \,g_R} \biggl\{g_R^2
\left[\left(\rho_{ud R}^A\right)^2+\left(\rho_{ud R}^V\right)^2\right]
-2
\left[\left(\rho^A_{udR}\right)^4+4 \left(\rho_{ud R}^A\right)^2
\left(\rho_{ud R}^V\right)^2+\left(\rho^V_{udR}\right)^4\right]\nonumber\\& - &
4 \left(\rho_{ud R}^A\right)^2
\left(\rho_{ud R}^V\right)^2 \cos (2 \,\theta^A_{udR}) \biggr\},
\label{deltag}
\end{eqnarray}
obtained from the direct calculation, has been taken into account.

   Equation (\ref{ConsCond2}) implies conditions on renormalized as well as on bare parameters. Demanding that the
renormalized couplings and counter-terms both comply with these conditions and taking into account Eq.~(\ref{ctts}) we obtain
two restrictions on couplings:
\begin{eqnarray}
&&g_R^4 \left[-\left(\left(\rho_{ud R}^A\right)^2+\left(\rho_{ud R}^V\right)^2\right)\right]
+8 \left(\rho_{ud R}^A\right)^2 \left(\rho_{ud R}^V\right)^2 \cos (2 \,\theta^A_{udR})
\left[g_R^2-12\left(\left(\rho_{ud R}^A\right)^2+\left(\rho_{ud R}^V\right)^2\right)\right]
\nonumber\\
&& +8g_R^2 \left[\left(\rho_{ud R}^A\right)^4
+5 \left(\rho_{ud R}^A\right)^2 \left(\rho_{ud R}^V\right)^2+\left(\rho_{ud R}^V\right)^4\right]\nonumber\\
&& -16 \biggl[\left(\rho_{ud R}^A\right)^6 +9 \left(\rho_{ud R}^A\right)^4 \left(\rho_{ud R}^V\right)^2
+9 \left(\rho_{ud R}^A\right)^2 \left(\rho_{ud R}^V\right)^4+\left(\rho_{ud R}^V\right)^6\biggr] =0,\nonumber\\
&&  5g_R^4 \left(\rho_{ud R}^A\right)^4+10g_R^4 \left(\rho_{ud R}^A\right)^2 \left(\rho_{ud R}^V\right)^2+5g_R^4
\left(\rho_{ud R}^V\right)^4 +16 \left(\rho_{ud R}^A\right)^2 \left(\rho_{ud R}^V\right)^2 \cos (2
\,\theta^A_{udR})\nonumber\\
&&\times \biggl[8 \left(3 \left(\rho_{ud R}^A\right)^4+8
\left(\rho_{ud R}^A\right)^2 \left(\rho_{ud R}^V\right)^2+3 \left(\rho_{ud R}^V\right)^4\right)
-3g_R^2 \left(\left(\rho_{ud R}^A\right)^2+\left(\rho_{ud R}^V\right)^2\right)\biggr]\nonumber\\
&&
-40g_R^2\left(\rho_{ud R}^A\right)^6 -232g_R^2 \left(\rho_{ud R}^A\right)^4
\left(\rho_{ud R}^V\right)^2-232g_R^2 \left(\rho_{ud R}^A\right)^2
\left(\rho_{ud R}^V\right)^4 -40g_R^2 \left(\rho_{ud R}^V\right)^6\nonumber\\
&&+32
\left(\rho_{ud R}^A\right)^4 \left(\rho_{ud R}^V\right)^4 \cos (4
\,\theta^A_{udR})+80 \left(\rho^A_{udR}\right)^8+768 \left(\rho_{ud R}^A\right)^6 \left(\rho_{ud R}^V\right)^2
+80 \left(\rho^V_{udR}\right)^8\nonumber\\
&& +1600\left(\rho_{ud R}^A\right)^4
\left(\rho_{ud R}^V\right)^4+768\left(\rho_{ud R}^A\right)^2\left(\rho_{ud R}^V\right)^6=0.
\label{twoc}
\end{eqnarray}
These two equations admit several solutions for $\rho_{ud R}^A$,  $\rho_{ud R}^R$ couplings and the angle $\theta^A_{udR}$,
which result in the following final expressions for the matrices of couplings:
\begin{eqnarray}
T_A^a & = & 0, \ \ T_V^a =0,\label{00}\\
T_A^a & = & 0, \ \ T_V^a =g\,\frac{\tau^a}{2} ,\label{0t}\\
T_A^a & = & 0, \ \ T_V^1 =-g\,\frac{\tau^1}{2}, \ \ T_V^2 =-g\,\frac{\tau^2}{2}, \ \ T_V^3 =g\,\frac{\tau^3}{2},\label{0tt}\\
T_A^1 & = & g\,\frac{\tau^1}{2}, \ \ T_A^2 =g\,\frac{\tau^2}{2}, \ \ T_A^3 =0, \ \ T_V^1 =0, \ \ T_V^2 =0, \ \ T_V^3 =g\,\frac{\tau^3}{2},\label{t0}\\
T_A^1 & = & -g\,\frac{\tau^1}{2}, \ \ T_A^2 =-g\,\frac{\tau^2}{2}, \ \ T_A^3 =0, \ \ T_V^1 =0, \ \ T_V^2 =0, \ \ T_V^3 =g\,\frac{\tau^3}{2},\label{tt0}\\
T_A^a & = & \pm g\,\frac{\tau^a}{4}, \ \ T_V^a = g\,\frac{\tau^a}{4}.
\label{CCsolutions}
\end{eqnarray}
By changing the overall phase of fermion fields the case (\ref{0tt}) reduces to (\ref{0t}). By defining the new fermion field $\psi_u'=\pm \gamma_5 \psi_u$ the cases (\ref{t0}) and (\ref{tt0}) (respectively to the sign in new fermion field) are reduced to (\ref{0t}). Further, by defining $\psi_{u,d}'=\gamma_5 \psi_{u,d}$ the first case in Eq.~(\ref{CCsolutions}), i.e. with positive sign, reduces to the second one, i.e. with negative sign. Thus for the most general interaction of vector bosons with a couple of massless fermions finally we obtain three possibilities
\begin{eqnarray}
T_A^a & = & 0, \ \ T_V^a =0,\label{00}\\
T_A^a & = & 0, \ \ T_V^a =g\,\frac{\tau^a}{2} ,\label{0t1}\\
T_A^a & = & - g\,\frac{\tau^a}{4}, \ \ T_V^a = g\,\frac{\tau^a}{4}.
\label{CCsolutionsFinal}
\end{eqnarray}
For an interacting parity non-conserving theory we are left with $V-A$ structure of the couplings specified by Eq.~(\ref{CCsolutionsFinal}).
Thus, finally we obtain an $SU(2)_L$ locally invariant Lagrangian
\begin{equation}
{\cal L}_{\rm F}  =  -{1\over 2} \ {\rm Tr} \left[G_{\mu\nu} G^{\mu\nu}\right]
+M^2\, {\rm Tr} \left[V_\mu V^{\mu}\right]
+ i\,\bar \Psi_L
D \hspace{-.6em}/\hspace{.1em}
\Psi_L+i\,\bar \Psi_R
\partial \hspace{-.5em}/\hspace{.1em}
\Psi_R ,
\label{EffLagr30}
\end{equation}
where
\begin{eqnarray}
D_\mu & = & \partial_\mu-i\,g V_\mu,\nonumber\\
\Psi_L & = & \frac{1-\gamma_5}{2}\,\Psi,\nonumber\\
\Psi_R & = & \frac{1+\gamma_5}{2}\,\Psi,\nonumber\\
V_\mu & = & t^a V^a_\mu,\ \ \
t^a =  \frac{\tau^a}{2},\nonumber\\
G_{\mu\nu} & = & \partial_\mu V_\nu -
\partial_\nu V_\mu - i\,g \left[V_\mu, V_\nu\right].
\label{gdefinition}
\end{eqnarray}
The local symmetry of the Lagrangian in Eq.~(\ref{EffLagr30}) is only broken by an $SU(2)$ globally invariant mass term of vector bosons.

\section{More fermions and renormalizability}

	It is extremely complicated to include more fermions in the analysis of the previous section, especially if the masses and mixing of the fermions are also taken into account. Instead, motivated by the obtained results
let us consider an EFT Lagrangian of massive Yang-Mills vector fields interacting with fermions given by
\begin{eqnarray}
{\cal L} & = & -{1\over 2} \ {\rm Tr} \left[G_{\mu\nu} G^{\mu\nu}\right]
+M^2\, {\rm Tr} \left[V_\mu V^{\mu}\right] +d_s \left[\partial_\mu s_{hf}
-s_{hf} V_\mu\right]\left[\partial^\mu s_{fh}^\dagger- V^{\mu} s_{fh}^\dagger\right] \nonumber\\
&+& i\,\bar \psi_L^f\,
D \hspace{-.6em}/\hspace{.1em}
\psi_L^f+i\,\bar \psi_R^h\,
\partial \hspace{-.6em}/\hspace{.1em}
\psi_R^h -\bar \psi_R^h\,s_{hf}
\psi_L^f 
- \bar \psi_L^f\,s_{fh}^\dagger
\psi_R^h
+ {\cal L}_1(\psi_R^h,\psi_L^f,B_\mu,s)\,,
\label{EffLagr3}
\end{eqnarray}
where the summation over $f$ runs from 1 to $N$, corresponding to left doublets, and the summation over $h$ runs from 1 to $2 N$ corresponding to right singlets, $s_{hf}$ is an external $N\times N$ matrix field and ${\cal L}_1$ stands for an infinite number of terms involving more fields and/or
derivatives, i.e. terms with dimension-full coupling constants. These terms are such that the Lagrangian of Eq.~(\ref{EffLagr3}), except the mass term of
the vector bosons, would be invariant under local transformations
\begin{eqnarray}
V_\mu(x) & \to & \Omega(x) V_\mu(x)\Omega^{-1}(x) + \Omega(x)\partial_\mu \Omega^{-1}(x)\,,\nonumber\\
\bar\psi_L^f(x) &\to&  \bar\psi_L^f(x)\Omega^{-1}(x)\,,\nonumber\\
\psi_L^f(x) &\to& \Omega(x) \psi_L^f(x)\,,\nonumber\\
\bar\psi_R^h(x) &\to&  \bar\psi_R^h(x)\,,\nonumber\\
\psi_R^h(x) &\to& \psi_R^h(x)\,,
\label{gaugetransfVmA1}
\end{eqnarray}
if we also transformed the external field matrix $s_{hf}$ in the following way
\begin{equation}
s_{hf} \to s_{hf}\Omega^{-1}(x).
\label{str}
\end{equation}
Note that the effective lagrangian also includes the derivatives of the $s_{hf}$ matrix, which vanish for the case of
constant $s_{hf}$ corresponding to the actual physical case, but give non-trivial contributions under transformations of Eq.~(\ref{str}).

The Lagrangian of Eq.~(\ref{EffLagr3}) defines a self-consistent renomalizable EFT which, for the appropriate choice of  $N$ and the external matrix field as the constant matrix of fermion masses and mixing parameters, coincides with the $SU(2)$ part of the WS model in unitary gauge with the scalar field put equal to its vacuum expectation value.

     To investigate the renormalizability of the EFT given by the Lagrangian of Eq.~(\ref{EffLagr3}), let us add a ''mass term''
of the external filed $\sigma^2 {\rm Tr}(s^\dagger s)$, where $\sigma$ is a constant parameter,  and consider $s_{hf}$ as a dynamical field.
Generating functional for Green's functions in the obtained massive Yang-Mills theory has the form
\begin{equation}
Z[J^a_\mu,I_s,\xi,\bar\xi] = \int {\cal D} s\,{\cal D} V\,{\cal D} \psi\,{\cal D}
\bar\psi\,e^{ i \int d^4 x \,\left[{\cal
L}(x)+J^{a\mu}V^a_\mu+I_s s +\xi_R \bar\psi_R + \bar \xi_R\psi_R
+\xi_L \bar\psi_L + \bar \xi_L\psi_L
\right]}\,.
\label{GFFinal2}
\end{equation}
    To investigate the symmetries of the effective action, following Refs.~\cite{'tHooft:1971fh,Slavnov:1972fg},
we make change of variables by performing an infinitesimal gauge transformation of in
Eq.~(\ref{GFFinal2})
\begin{eqnarray}
B^a_\mu(x) & \to & V^a_\mu(x) + g\,f^{abc} V_\mu^b(x)
\phi^c(x)+\partial_\mu \phi^a(x)\,,\nonumber\\
\bar\psi_L^f(x) &\to&  \bar\psi_L^f(x)-i\,g\,\bar\psi_L^f(x)\,
t^a\, \phi^a \,,\nonumber\\
\psi_L^f(x) &\to& \psi_L^f(x)+i\,g\,t^a
\,\psi_L^f(x)\,\phi^a(x)\,,\nonumber\\
\bar\psi_R^h(x) &\to&  \bar\psi^h_R(x)\,,\nonumber\\
\psi_R^h(x) &\to& \psi_R^h(x)\,,\nonumber\\
s_{hf}(x) &\to& s_{hf}(x)-i\,g\,s_{hf}(x)\,
t^a\, \phi^a,
\label{gaugetransfVmA}
\end{eqnarray}
where $\phi^a(x)$ are arbitrary infinitesimal functions.

    The change of variables does not change the generating functional
$Z[J^a_\mu,\xi,\bar\xi]$.
The measure is invariant under transformations of Eq.~(\ref{gaugetransfVmA}) up to the anomalous terms which cancel each other if the fermion content of the effective Lagrangian is the same as in the SW model, i.e. the equal number of quark and lepton generations are included with right quantum numbers. The action in Eq.~(\ref{GFFinal2}) is also invariant up to the mass term of vector
bosons and the terms involving external sources $J_\mu$, $I_s$, $\xi_L$ and $\bar\xi_L$. Thus we obtain
\begin{eqnarray}
&& \int {\cal D} s\,{\cal D} V\,{\cal D} \psi\,{\cal D} \bar\psi\,\exp i \left\{
\int d^4 x \,\left[ {\cal L}(x)+J^{a\mu}V^a_\mu+I_s s+\xi_R \bar\psi_R + \bar \xi_R\psi_R
+\xi_L \bar\psi_L + \bar \xi_L\psi_L\right]\right\} \nonumber\\
&& \times
 \int d^4 z \biggl[ - M^2 V^{c \mu}(z)\partial_\mu \phi^c(z)-
J^{c \mu}(z) \partial_\mu\phi^c(z)- g\,f^{abc} J^{a \mu}(z)\,V^b_\mu(z)\phi^c(z)\nonumber\\
&  & +i\,g\,I_s(z)\,s(z)\,t^c \phi^c(z)+i\,g\,\xi_L(z)\,\bar\psi_L(z)\,t^c \phi^c(z)- i\,g\,\bar \xi_L(z)\,t^c
\psi_L(z)\phi^c(z) \biggr]
=0\,.
\label{STIVmA}
\end{eqnarray}
	Equation (\ref{STIVmA}) can be rewritten as
\begin{eqnarray}
&& \int d^4 z \biggl[
\frac{\delta\Gamma[V,s,\bar\psi_{L,R},\psi_{L,R}]}{\delta V^c_\mu(z)} \partial_\mu\phi^c(z) +g\,f^{abc} \frac{\delta\Gamma[V,s,\bar\psi_{L,R},\psi_{L,R}]}{\delta V^a_\mu(z)}\,<V^b_\mu(z)>_S\phi^c(z)\nonumber\\
&& -i\,g\,\frac{\delta\Gamma[V,s,\bar\psi_{L,R},\psi_{L,R}]}{\delta s}<s(z)>_S\,t^c \phi^c(z)-i\,g\,\frac{\delta\Gamma[V,s,\bar\psi_{L,R},\psi_{L,R}]}{\delta \bar\psi_L(z)}<\bar\psi(z)>_S\,t^c \phi^c(z)\nonumber\\
&&+i\,g\,\frac{\delta\Gamma[V,s,\bar\psi_{L,R},\psi_{L,R}]}{\delta\psi_L(z)} \,t^c<\psi(z)>_S \phi^c(z)- M^2 <V^{c \mu}(z)>_{S}\partial_\mu \phi^c(z) \biggr]=0, \label{IdentityRewritten}
\end{eqnarray}
where
\begin{eqnarray}
Z[J,I,\bar\xi,\xi] < O(z) >_S &=&
\int {\cal D} s\,{\cal D} V\,{\cal D} \psi\,{\cal D} \bar\psi\,O(z)\,\exp i \Biggl\{
\int d^4 x \,\biggl[ {\cal L}(x)+J^{a\mu}V^a_\mu \nonumber\\
&+& I_s s+\xi_R \bar\psi_R + \bar \xi_R\psi_R
+\xi_L \bar\psi_L + \bar \xi_L\psi_L\biggr]\Biggr\}
\label{<O>definition}
\end{eqnarray}
and $\Gamma[V,s,\bar\psi_{L,R},\psi_{L,R}]$ is the effective action.

It follows from Eq.~(\ref{IdentityRewritten}) that the variation of the effective action under gauge
transformations is equal to the variation of the globally invariant mass term of vector bosons.
Hence the effective action is gauge invariant if we drop the mass term, exactly as for the effective Lagrangian. Divergences of all loop diagrams generated by a local Lagrangian of any quantum field theory
can be subtracted systematically by using the Zimmerman's forest formula. These subtractions can be generated by including corresponding local counter terms in the Lagrangian (see e.g. Ref.~\cite{Collins:1984xc}). Due to the symmetries of the effective action all counter terms in our effective field theoretical
model satisfy the constraints of gauge invariance. As the effective Lagrangian contains {\it all} terms which are invariant under gauge transformations, all these counter terms can be absorbed by redefining corresponding parameters and fields of the effective Lagrangian  \cite{Weinberg:mt}.

We are not interested in a theory with dynamical $s(x)$  fields. However, the $N$-point Green's functions of vector bosons and fermions of our EFT model with constant external field $s$ correspond to the (multiple) sum of the $N+k$-point Green's functions of the EFT with dynamical $s(x)$.  All of these $N+k$-point Green's functions do not contain internal lines corresponding to $s(x)$ field and the first $N$ legs correspond to the same vector boson and fermion fields as in the $N$-point Green's functions which we are interested in. The rest $k$  legs correspond to the $s(x)$ field with vanishing momenta for these legs. From these considerations we conclude that our EFT model Lagrangian which is obtained from Eq.~(\ref{EffLagr3}) by replacing the dynamical $s(x)$ by the constant matrix is renormalizable in the sense of EFT.

Constant matrix $s$ can be diagonalized in a standard way, leading to the masses of fermions and the Cabibbo-Kobayashi-Maskava mixing matrix \cite{Cabibbo:1963yz,Kobayashi:1973fv}. To include the electromagnetic interaction, we need to add a massless vector field by demanding an exact $U(1)$ local gauge invariance of the resulting theory. This is because it is assumed that the photons are exactly massless. The physical photon and the $Z$-boson fields are obtained by mixing the additional $U(1)$ vector field and $V_\mu^3$  \cite{Djukanovic:2005ag}. Leading order of the final effective Lagrangian coincides with the Weinberg-Salam model in which the scalar field is replaced by its vacuum expectation value.

\section{summary and discussions}

      This work is an attempt to probe the following question: based on the modern understanding of quantum field theories what kind of theory
of weak interaction would we construct if it did not exist yet?

Weak interaction is mediated by massive vector bosons. While a Lorentz invariant local QFT of massles vector particles has to be (locally) gauge invariant, for massive vector bosons there is no reason to assume local gauge symmetry. Renormalizability in the traditional sense, which lead to a theory of electro-weak interactions with a spontaneously broken gauge symmetry, is replaced by renormalizybility in the sense of effective field theories.  However this model describes the experimental data with impressive accuracy. Is there any explanation for this (seeming?) logical mismatch? These considerations motivate to re-examine the underlying principles of the Weinberg-Salam model.

	In trying to probe the problem formulated above we analyzed the most general Lorentz invariant EFT of massive vector bosons.
 The effective Lagrangian contains an infinite number of interaction
terms. We assumed that all interactions with couplings of
negative mass-dimensions are suppressed by powers of some scale which is much larger than the characteristic energy scale of problems to which we apply the considered EFT. Massive vector bosons are spin one particles and therefore they have to be described by Lagrangians with
constraints. This implies that the coupling constants of the effective Lagrangian satisfy some non-trivial relations.
Furthermore, demanding the perturbative renormalizability in the sense of
EFT additional consistency conditions are imposed.
As a result we are left with $SU(2)$ {\it locally} invariant Lagrangian up
to the globally invariant mass term of vector bosons.
	Next we introduced a doublet of massless fermions in the above EFT and demanded
the perturbative renormalizability in the sense of EFT.  As a result we obtained a leading order effective Lagrangian which is locally invariant up to the globally invariant vector boson mass term. The symmetry group is
either $SU(2)$, i.e. vector bosons interact with vector current, or $SU(2)_L$, i.e. vector bosons interact with vector minus axial vector currents. Parity non-conservation suggests that for the weak interaction the $SU(2)_L$ locally invariant Lagrangian should be chosen. As a self-consistent UV completion of the obtained leading order effective Lagrangian we considered an $SU(2)_L$ locally invariant EFT and
introduced the fermion masses and mixing by considering them as external fields. By including in the effective Lagrangian all terms which are invariant under local transformations, when the external fields are also transformed, and after choosing the external fields to be equal to the constant fermion mass matrix we obtained an perturbatively renormalizable EFT.
The electromagnetic interaction can be included in the effective Lagrangian  following Ref.~\cite{Djukanovic:2005ag} taking into account that as the photons are massless, the corresponding effective Lagrangian has to be $U(1)$ gauge invariant. The leading order part of the final effective Lagrangian coincides with the WS model in unitary gauge with the scalar field put equal to its constant vacuum expectation value.

	Perturbative unitarity is often appealed as a strong argument against the theories of massive
vector bosons without spontaneous symmetry breaking. While the perturbation theory is a powerful tool in quantum field theories, some conclusions drawn from
perturbative arguments might be completely misleading. For example, if there was an exact contribution $g/(1+g^2 s/M^2)$ in some physical amplitude/cross section, it would vanish for large $s$, while the perturbative expansion $g-g^3 s/M^2+\cdots$ suggests that given physical quantity diverges strongly for large $s$. 
At sufficiently high energies
the perturbative unitarity is broken in any EFT, including the standard model treated as a leading order approximation to the corresponding EFT.
Note that the breakdown of the perturbative unitarity
does not necessarily mean that new degrees of freedom should be included in the theory at breakdown scale. For example, this could be an indication that resonant states appear.
Another possible scenario of the restoration of unitarity could be provided by classicalization \cite{Dvali:2010jz,Dvali:2010ns,Dvali:2011nj}.

To conclude, the Weinberg-Salam model does such a good job in describing the available experimental data because the most general self-consistent (parity non-conserving) leading order EFT Lagrangian of massive vector bosons interacting with fermions and electromagnetic field looks like as if it was an
$SU(2)_L\times U(1)$ gauge invariant theory with spontaneous symmetry breaking taken in unitary gauge and the scalar field substituted with its vacuum expectation value. The results of the present work imply that it is natural to expect that the Higgs particle will not be discovered despite the preliminary claims that the mass $\sim 125$ GeV is favored by LHC data \cite{ATLAS:2012ae,Chatrchyan:2012tx,ATLAS:2012ad,Chatrchyan:2012tw}.

One might object: even if the logic of the current work is correct it certainly contradicts the various precision experimental tests of the standard model. However it does not! The fact that the Weinberg-Salam model agrees with available experimental data means that the couplings of ''non-renormalizable'' interactions are suppressed by some large scale and their effects will show up only at sufficiently high energies (the anomalous magnetic moment of the muon might be the first messenger of such terms). This is similar to QED - despite its extremely good accuracy in describing the experimental data, QED is only the leading order approximation to an EFT in which the coupling constants of  higher-order ''non-renormalizable' terms are suppressed by some very large scale \cite{Weinberg:mt}.


\acknowledgments

This work was supported by the Deutsche Forschungsgemeinschaft (grant Nr. GE 2218/2-1)
and Georgian Shota Rustaveli National Science
Foundation (grant Nr. 11/31).


\end{document}